\documentstyle[twoside,fleqn,espcrc2,epsf]{article}

\newcommand{\beq}{\begin{equation}}
\newcommand{\eeq}{\end{equation}}

\newcommand{\AmS}{{\protect\the\textfont2
  A\kern-.1667em\lower.5ex\hbox{M}\kern-.125emS}}

\title{Simplicial Gravity In Dimension Greater Than Two}

\author{S. Catterall$^{\rm a}$, G. Thorleifsson
        \address{Physics Department, Syracuse University,
        Syracuse, NY 13244}, 
        J. Kogut\address{Loomis laboratory of Physics,
        University of Illinois, Urbana, IL 61801}
        and
        R. Renken\address{Physics Department, University of Central
        Florida, Orlando, FL 32816}} 
       
\begin{document}

\begin{abstract}
We consider two issues in the DT model of quantum gravity. 
First, it is shown that the
triangulation space for $D>3$ is dominated by 
triangulations containing a single {\it singular} $(D-3)$-simplex 
composed of vertices with divergent dual volumes. 
Second we study the
ergodicity of current simulation algorithms. Results from runs
conducted close to the 
phase 
transition of the
four-dimensional theory are shown. We see no strong indications of ergodicity breaking 
in the simulation and our data support recent claims that the transition
is most probably first order. Furthermore, we show that the critical
properties of the system are determined by the dynamics of remnant
singular vertices. 
\end{abstract}

\maketitle

\section{Quantum Gravity}

The problem of reconciling the two theories of general relativity and
quantum mechanics is perhaps one of the most important and difficult
issues in theoretical physics. A variety of approaches have been
adopted see eg.\cite{rev} but it is fair to say that only limited success
has been achieved so far. In this article we shall concern
ourselves with one such approach in which quantum mechanical fluctuations
of the (Euclidean) geometry are encoded in a path integral. 

\beq
Z\left(G\right)=\int {\cal D}[g] \exp{\left[-S\left(g,G\right)\right]}
\label{eqn1}
\eeq

We can take the action to be the standard Einstein-Hilbert
term $S\left(g\right)=\frac{1}{16\pi G}\int d^4x\sqrt{g}\,R$ with gravitational coupling
$G$. A variety of difficulties are immediately observed; firstly, $G$
carries dimensions of inverse mass squared so that the model is
{\it perturbatively} nonrenormalizable. Perhaps worse $S$ is unbounded
from below so it is not even clear that a stable vacuum exists. 
Furthermore, even
when the functional integral is
restricted  to some fixed topology it is not clear that a unique choice
exists for the functional measure ${\cal D}[g]$. The approach we have been
following is to replace this continuum integral with a finite lattice sum
in which these problems can be evaded or at least regulated. This
approach is termed DT-gravity. Some of the excitement about this
model has stemmed from the initial observations \cite{initial} that
the four-dimensional theory appeared to exhibit a phase transition at
finite bare $G_0$. This opened up the possibility of defining
a continuum theory in the vicinity of this nonperturbative fixed point, 
perhaps in the spirit of Weinberg's work on $2+\epsilon$ gravity 
\cite{wein}.
  
\section{Dynamical Triangulations}

The basic idea is that one can divide $D$-dimensional space into a collection
of equal volume elementary cells. The simplest choice for a cell is that
it is a $D$-simplex --- a set of $(D+1)$ fully interconnected points with
edge length $a$ taken to be some invariant cut-off. A piecewise linear
approximation to the manifold is then constructed by gluing these simplices
together across their $(D-1)$-dimensional faces. Additional restrictions
are also commonly employed in order to ensure that the neighborhood of any
vertex is homeomorphic to a $D$-ball. The local scalar curvature density
is then proportional to the number of simplices common to a given
$(D-2)$-simplex which is bounded if the total volume is held fixed.
Furthermore, its integral, the lattice action is essentially just the
total number of nodes in the triangulation.

The partition function eqn~\ref{eqn1} is then replaced by the lattice sum

\beq
Z=\sum_T\exp{\left[\kappa_0 N_0 -\kappa_4 N_4
-\left(\frac{\left(N_4-V\right)}{\Delta V}\right)^2\right]}
\eeq

In practice the bare cosmological constant $\kappa_4$ is tuned to
achieve a mean volume $\left<N_4\right>=V$ and the remaining coupling
$\kappa_0$ plays the role of a bare Newton constant $G_0$. The parameter
$\Delta V$ controls the size of the volume fluctuations. 

There are many unknowns in this model. In this paper we consider just
two --- the structure of the measure over triangulations and the
ergodicity of the simulation algorithms currently in use. As we shall see there are
some intimate connections between these. 
 
\section{Singular Structures}

Consider the triangulation space of the fixed volume $V$ ensemble
for $D>3$. Let us define the local volume $\omega_i$ of some
(sub)$i$-simplex as the number of simplices which contain that
$i$-simplex. Let us also call this simplex {\it singular} if 
$\omega_i\to\infty$ as $V\to\infty$.

Our results can be summarized in two conjectures:

\begin{itemize}
\item Every triangulation contains {\it exactly one} singular
$(D-3)$-simplex composed of $(D-2)$ singular vertices.
\item The local volume $\omega\left(D-3\right)\sim aV^{\frac{2}{3}}$ while
the singular vertex local volumes $\omega_0\sim bV$
\end{itemize}

\begin{figure}[htb]
\vspace{9pt}
\epsfxsize=2.8 in
\epsfbox{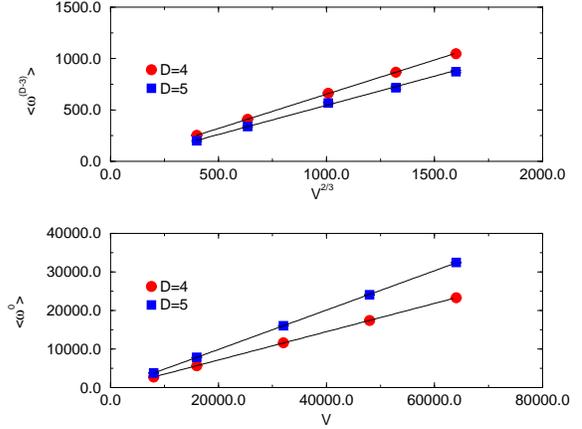}
\caption{Singular local volumes for D=4,5}
\label{fig1}
\end{figure}

Fig~\ref{fig1} consists of two plots which illustrate these facts. The
upper plot shows the volume of the singular $(D-3)$-simplex for $D=4,5$
against the two-thirds power of the total volume $V$. Clearly the data
lies on a straight line verifying the first half of the second
conjecture. The lower plot shows that the singular vertex volumes 
vary linearly with the total volume confirming the second
part of this conjecture. Indeed, we have observed similar results
in dimensions $D=6,7$ and can state that {\it all} the descendent
subsimplices of the original singular $(D-3)$-simplex have local volumes
which diverge {\it linearly} with the total volume in contrast to the
{\it sublinear} growth observed for the $(D-3)$-simplex. Simple
geometrical arguments are given in \cite{us} for the two thirds power in
the latter case. Indeed, the coefficient $a$ can also be predicted in
terms of $b$ in this model and appears to fit the data rather well.

The immediate question that arises is why there are so many 
triangulations possessing this singular structure. It is simple
to understand this by introducing the concept of a {\it local}
entropy. Consider the set of simplices which form the local volume
$\omega_i$ of some $i$-simplex and remove the common $(i+1)$ vertices.
The remaining vertices and their interconnections then form a
triangulation of a {\it dual} $(D-1-i)$ sphere. For example, a circle
$S^1$ forms the boundary of the dual face to a link in three dimensions.
The local entropy $s_i$ ascribed to that $i$-simplex is then defined to
be equal to the number of ways of assembling the simplices in the
local volume or equivalently {\it the number of triangulations} of
this dual $(D-1-i)$-sphere.

\beq
s_i=Z_{D-1-i}\left(\omega_i\right)
\eeq

This function is known to increase {\it at least} exponentially fast with
volume for spheres with dimension greater than unity. Thus, $i$-simplices
with $(D-1-i)\ge 2$ i.e.\ $i\le (D-3)$ can increase their local entropy by 
growing their local volumes. It is then intuitively reasonable that
a single singular $(D-3)$-simplex ultimately dominates composed of a
cascade of smaller dimension secondary simplices. 

\begin{figure}[htb]
\vspace{9pt}
\epsfxsize=2.8 in
\epsfbox{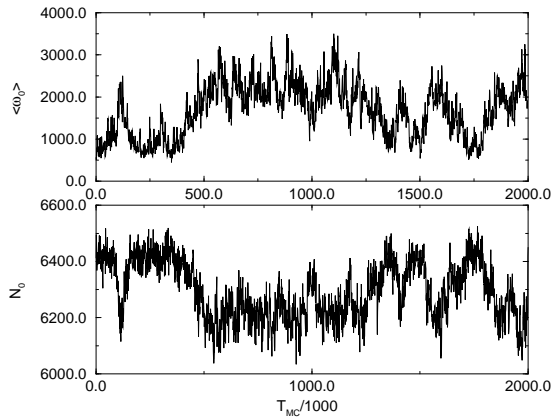}
\caption{Correlation of singular vertex volume with total vertex number
close to the transition}
\label{fig2}
\end{figure}

This structure then forms a nonperturbative background in the
crumpled phase of the model around which
small fluctuations occur. Preliminary studies in four
dimensions show that the volume
of these singular vertices decreases monotonically as the coupling
$\kappa_0$ increases. This
is eminently reasonable since forcing more vertices into
the system by increasing the coupling enhances
the contribution of less singular triangulations (Singular ones have
a minimal number of vertices for fixed volume).

Close to the phase transition a gas of remnant
singular vertices exists whose fluctuations are highly correlated with
the fluctuations of the total number of vertices. This is
illustrated in fig~\ref{fig2} which shows a perfect (anti)correlation
between the singular vertex volume $\omega_0$ and total
vertex number $N_0$ for $V=32K$, $D=4$, $\Delta V=100$ and $\kappa_0=2.516$ (the
pseudocritical coupling). Thus
the peak in the node susceptibility is driven by the fluctuations
in the singular simplex structure. The state with $\langle N_0 \rangle\sim 6400$ has
a large mean extent and essentially no singular vertices --- a local
volume of approximately 1000 means that the vertex has
merged with the background distribution of vertex local volumes and
can be no longer distinguished as a {\it singular} vertex. Correspondingly
the state with $\langle N_0\rangle\sim 6200$ has a small extent and contains a 
variable number (at least one) of remnant singular vertices which
are well separated from the background distribution. The critical
system appears to tunnel repeatedly between these two states. We will
return to this issue when we describe our experiments to check
ergodicity.

\section{Ergodicity}

Current simulation algorithms typically target some volume $V$ and allow
for some (usually rather small) fluctuations of size $\Delta V$ around this
target value. It is known that for $\Delta V\to\infty$ the Monte Carlo
algorithm used in the simulation is ergodic. That is all triangulations
may be reached by the application of the elementary moves. However, the
ergodic properties of the algorithm for finite $\Delta V$ are unknown.
Indeed for $\Delta V< D$ the moves are known to be {\it not}
ergodic. Thus it is important to check ergodicity carefully when studying
the theory at finite $\Delta V$. In the light of recent claims that
the four-dimensional theory has only a first order transition we have
studied that model close to criticality, using large lattices and for
three different values of the fluctuation volume $\Delta V=10,100,1000$.
A dependence of expectation values on $\Delta V$ would signal a 
breaking of ergodicity. One way in which this could occur would be
to suppose that the paths in the triangulation space between two
volume $V$ triangulations required intermediate triangulations with
much larger volumes. Barriers of height $B\left(V\right)$ might exist
between states. Ergodicity would then be broken if

\beq
\Delta V \ll B(V)
\eeq

\begin{figure}[htb]
\vspace{9pt}
\epsfxsize=2.8 in
\epsfbox{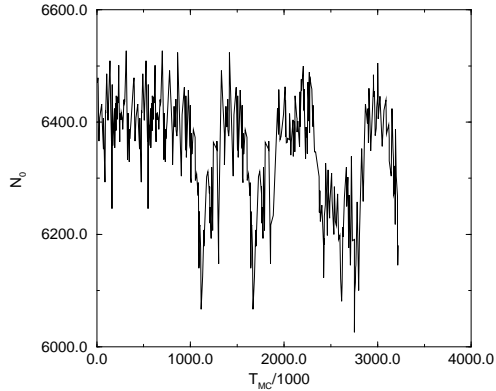}
\caption{Time series for $V=32K$, $\Delta V=1000$, $\kappa_0=2.516$}
\label{fig3a}
\end{figure}

Fig~\ref{fig3a} shows data for $V=32K$ and $\kappa_0=2.516\sim \kappa_0^c$
for $\Delta V=1000$. As for the earlier data presented
in fig~\ref{fig2} the Monte Carlo time series clearly shows a sequence
of tunneling events between two metastable states - this is the origin of
the first order signal reported in \cite{biel}. Similar signals are seen
at $\Delta V=10$ --- we see no sign of a dependence of expectation
values on $\Delta V$. Again, the two states can be labeled
by a zero and non-zero number of remnant singular vertices. 

\begin{figure}[htb]
\vspace{9pt}
\epsfxsize=2.8 in
\epsfbox{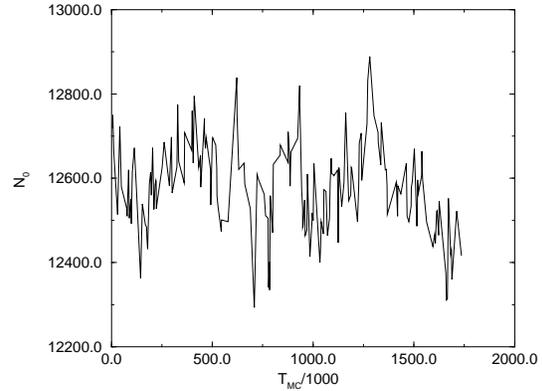}
\caption{Time series for $V=64K$, $\Delta V=1000$, $\kappa_0=2.56$}
\label{fig5}
\end{figure}

Unfortunately, the situation is less clear for $V=64K$. A possible
two-state signal is observed for $\Delta V=10,100$. The two
states correspond to $\langle N_0\rangle\sim 12600$ and $\langle N_0\rangle\sim
13000$. However, since
only one tunneling event is observed it cannot truthfully
be distinguished from transient behavior associated with
equilibration. Indeed, fig~\ref{fig5}
shows the Monte Carlo time series for $\Delta V=1000$ and
$\kappa_0=2.56\sim\kappa_0^c$ in which only the $\langle N_0\rangle=12600$
state is seen. 
After more than two million sweeps
there is still no sign of a tunneling event to the other state. Clearly, it is difficult to
use this large volume data to infer very much about the order of
the transition due to the current lack of
statistics. 

\end{document}